\documentstyle[preprint,aps]{revtex}
\tightenlines
\draft
\begin{document}
\title{The Hydrodynamic Equations 
of Superfluid Mixtures in Magnetic Traps}

\author{Tin-Lun Ho and V.B. Shenoy}
\address{Physics Department,  The Ohio State University, Columbus, Ohio
43210}
\date{\today}
\maketitle

\begin{abstract} 
The hydrodynamics equations of binary mixtures of Bose gases, Fermi gases, 
and mixtures of Bose and Fermi gases in the presence of external potentials
are derived. These equations can be applied to current experiments on 
mixtures of atomic gases confined in magnetic traps. 
\end{abstract}

The recent discoveries of Bose-Einstein condensation of atomic
gases\cite{Colorado,MIT,Rice}
have generated much interest in studying dilute Bose gases trapped in
external
potentials. While most of the experiments are performed on trapped atoms of the 
same kind, 
 the current traps can also hold different kinds of atoms at 
the same time, providing opportunities to study mixtures of
Bose gases, mixtures of Fermi gases and  mixtures of Bose and Fermi gases. 
In a recent paper, we have shown\cite{HSmixture} 
that there is a great variety of ground state structures in the binary mixtures of
Bose gases. These structures can be experimentally accessed by varying the number of trapped Bosons of each kind. The two 
Bosons can be different alkalis, different isotopes of the same alkali, or 
different hyperfine spin ($F$) states of the same alkali isotope. 
Recently, a mixture of different hyperfine states of $^{87}$Rb ($F=1$ and $F=2$) 
has been produced by the JILA group\cite{JILAmixture}, creating a system of 
interpenetrating Bose superfluids for the first time. 

In the case of superfluid $^{4}$He, its superfluid properties are 
contained in the well known two-fluid hydrodynamic equations of
Landau\cite{Landau} 
that  incorporate the effect of broken gauge symmetry into  
hydrodynamics.  In order to understand the properties of mixtures, 
it is important to derive the corresponding hydrodynamic equations, 
which is the goal of this paper.  Our derivation is a standard one. 
It follows from general principles such as conservation laws,  
broken symmetries, thermodynamic relations and the second law of 
thermodynamics\cite{Landau,HoHe3}. 
It will be clear from the derivation that the form of the 
hydrodynamic equations is independent of the density of the gases, as long as
the symmetry of the system remains unchanged. What distinguishes a dilute and 
a dense system is the explicit form of the coefficients entering  
these equations, which are related to the thermodynamic potential of the
system.
In this paper, we will focus only on the derivation of the hydrodynamic
equations of mixtures. The solutions of these equations require the
evaluation of the thermodynamic potentials which will be discussed
elsewhere. (We have recently solved  the hydrodynamic
equations of single component dilute Bose gases in spherical traps
and have obtained all the first and second sound modes (classified by radial
and angular quantum numbers) as a function of temperature\cite{VHsound}). 

Before proceeding, it is necessary to comment on the broken symmetry of the
alkali gases in magnetic traps. Since the alkali atoms carry spin, the
order parameter in the Bose
condensed phase should be a spinor field instead of a scalar field. However, in a
magnetic trap, only the spin states locally aligned with the magnetic field
${\bf B}({\bf r})$ are trapped\cite{HSspin}.  The resulting quantum gas 
therefore has the  spin field ``frozen" along the local field, 
leaving particle motion as the only
degree of freedom. As a result, the spin field reduces to a scalar.
On the other hand, even though the spin field is frozen, the spinor nature 
is not completely erased. In a recent paper\cite{HSspin}, we have shown
that the spatial variations of the spin field will give rise to a Berry phase
which can lead to a number of dramatic effects.  While the Berry
phase effects in current traps are small, they can be magnified
by changing the geometry of the trap. For simplicity, we shall only 
consider those traps where the Berry phase effects are negligible. One should, 
however, keep in mind that the hydrodynamics equations derived here are 
limited to traps of this kind. 

Since both scalar Bose superfluids and s-wave Fermion superfluids have
identical broken $U(1)$ gauge symmetry, and since the hydrodynamics of
a system depends entirely on the nature of its broken symmetry, the 
hydrodynamic equations of these two systems are identical. 
The hydrodynamics of binary mixtures of quantum gases (which can be either 
Fermions or Bosons) therefore fall in three categories, namely, mixtures of 
superfluids, mixtures consisting of a normal fluid and a superfluid, and mixtures of 
normal fluids. The superfluids can be the scalar condensates of Bosons or 
the s-wave BCS condensates of Fermions. The normal fluid can be the 
either Bosons or Fermions in their normal state. 

Among all the cases considered, the hydrodynamics of binary superfluid mixtures 
in a trap is the most complex. We shall focus on this case. As we shall see, 
once we derive the hydrodynamic equations for this system,
extensions to other mixtures is easy. For a binary mixture of Bosons,
in general, there will be regions in the trap where the two species
coexist and regions where there is no coexistence. 
Furthermore, in the region of coexistence, each Boson 
species may be in their normal or superfluid states. 
Since the hydrodynamic modes
in each of different regions are connected to one another, 
not only does one need to know
the hydrodynamic equations of interpenetrating superfluids, but also those 
of the surrounding mixtures of normal and superfluids. 

The organization of this paper is as follows. In Section {\bf I}, 
we will discuss 
the hydrodynamic variables of binary mixtures of scalar superfluids and their
Galilean transformation properties. The hydrodynamic equations for this
system of interpenetrating scalar superfluids will be
derived in Section {\bf II}. 
The explicit form of the hydrodynamic equations (with and without 
dissipation) will  be presented. 
Section {\bf III} is a summary of the hydrodynamic equations for other mixture 
systems. 

\vspace{0.2in}

\noindent {\bf I :  The hydrodynamic variables of a binary superfluid mixtures,
conservation laws, and their Galilean transformation properties :} 

Before studying hydrodynamics, it is necessary to study the
equilibrium thermodynamics of homogeneous systems, which is described by 
the entropy function $S$. $S$ is a function of a set of thermodynamic
variables $\{ X_{i} \}$, which are quantities specifying the physical 
state of the system, $S=S(\{ X_{i} \})$. To study binary mixtures, it is useful
to consider the entropy functions of the following sequence of systems with 
increasing complexity : 

\noindent (i)  Single component normal fluids : 
In this case, $\{ X_{i}\} = (E, {\bf P}, N, V)$,  
$S=S(E, {\bf P}, N, V)$, where $E, {\bf P}, N$ and $V$ are the total energy, 
momentum, particle number and volume of the system respectively. 
The meaning of $E, {\bf P}$ and $V$ will remain unchanged in the following 
examples (ii) to (v). 

\noindent (ii) Binary mixtures of normal fluids : Each component in the
mixture can be either a Boson or a Fermion. 
In this case $S=S(E, {\bf P}, N_{1}, N_{2}, V)$, where $N_{1}$ and $N_{2}$
are the particle numbers of the two normal fluids labelled as {\bf 1} and 
{\bf 2} respectively. 
The reason that $S$ is a function of $N_{1}$ and $N_{2}$ but not a function
separate momenta ${\bf P}_{1}$ and ${\bf P}_{2}$ is 
because while Boson {\bf 1} does not turn into Boson {\bf 2}, 
they can exchange momenta through particle collisions. As a result, 
the equilibrium state of the system is  characterised by a single momentum.

\noindent (iii) Single component scalar superfluid : This is the well known 
case of $^{4}$He. The system now acquires a broken gauge symmetry 
$<\hat\psi({\bf r})>= |<\hat{\psi}({\bf r})>|e^{i\theta({\bf r})}$, where 
$\hat{\psi}$ is the field operator. In this case, 
$S=S(E, {\bf P}, N, V; {\bf v}_{s})$, where 
${\bf v}_{s}= (\hbar/m){\bf \nabla}\theta$ is the superfluid velocity
describing the phase winding in the system, 
and $m$ is the mass of the Boson. 

\noindent (iv) Binary mixtures consisting of a scalar superfluid and a
normal fluid : Denoting the 
superfluid and the normal fluid components 
as {\bf 1} and {\bf 2} respectively, we have for this case
$S=S(E, {\bf P}, N_{1}, N_{2}, V; {\bf v}_{s1})$. 
The reason that $S$ is a function of total momentum ${\bf P}$ instead of 
separate momenta ${\bf P}_{1}$ and ${\bf P}_{2}$ is the same as that for 
normal fluid mixtures. 

\noindent (v) Binary mixtures of superfluids : We denote the two superfluids 
as 1 and 2, and their masses by $m_{1}$ and $m_{2}$ respectively. 
From the discussion in cases (i) to (iv) above, we have 
\begin{equation}
S=S(E, {\bf P}, N_{1}, N_{2}, V; {\bf v}_{s1}, {\bf v}_{s2}),
\label{totalS} \end{equation}
with
\begin{equation}
{\bf v}_{s\alpha} \equiv \frac{\hbar}{m_{\alpha}}{\bf \nabla}\theta_{\alpha}, \,\,\,\,\,
\,\,\,\, \alpha = 1,2 . 
\end{equation}
From this point onwards, we shall consider only case (v). The hydrodynamic equations 
of cases (i) to (iv) can be obtained from the equations derived for case (v) by setting the
appropriate superfluid densities (defined below) to zero. (See section {\bf
(III)}).

The extensive property of $S$ implies 
\begin{equation}
S(E, {\bf P}, N_{1}, N_{2}, V; {\bf v}_{s1}, {\bf v}_{s2}) =
Vs(\epsilon, {\bf g}, \rho_{1}, \rho_{2}, {\bf v}_{s1}, {\bf
v}_{s2}),    \label{extensive}
\end{equation}
where $s=S/V$, $\epsilon=E/V$, ${\bf g}={\bf P}/V$ are the entropy density, 
the energy density,  
 the momentum density, and 
$\rho_{\alpha}=m_{\alpha}N_{\alpha}/V$ is the mass density of the species
labelled
$\alpha$ in the mixture, $(\alpha=1,2)$. 
The arguments $\{ \epsilon, {\bf g}, \rho_{\alpha}, {\bf v}_{s\alpha}\}$ 
of the entropy density $s$ will be referred to as the 
"hydrodynamic variables" of the system, $\{x_i\}$. 
Defining temperature $T$, chemical potential (per unit mass) 
$\mu_{\alpha}$, ``normal fluid velocity" ${\bf v}_{n}$, and the vector 
${\bf b}_{\alpha}$
as $\partial s/\partial\epsilon\equiv 1/T, 
\partial s/\partial\rho_{\alpha}\equiv -\mu_{\alpha}/T, 
\partial s/\partial{\bf g} \equiv -{\bf v}_{n}/T$, 
and $\partial s/\partial{\bf v}_{s\alpha} \equiv -{\bf b}_{\alpha}/T$
respectively, 
we can write
\begin{equation}
{\rm d}s = \frac{1}{T}{\rm d}\epsilon
-\frac{{\bf v}_{n}}{T}\cdot {\rm d}{\bf g} 
-\sum_{\alpha=1}^{2}\frac{\mu_{\alpha}}{T}{\rm d}\rho_{\alpha}
-\sum_{\alpha=1}^{2}\frac{{\bf b}_{\alpha}}{T}\cdot {\rm d}{\bf v}_{s\alpha}. 
\label{ds} \end{equation}
Defining the pressure as $P = T(\partial S/\partial V)_{E, {\bf P}, N}$, 
and using the extensivity relation given by Eq.(\ref{extensive}), we have
\begin{equation}
P =  Ts - \epsilon +\mu_{1}\rho_{1} + \mu_{2}\rho_{2} + 
{\bf v}_{n}\cdot{\bf g},   \label{pressure}
\end{equation}
and the Gibbs-Duham relation 
\begin{equation}
{\rm d} P =  s{\rm d} T + \sum_{\alpha=1}^{2}\rho_{\alpha}{\rm d} 
\mu_{\alpha}  +
{\bf g}\cdot{\rm d} {\bf v}_{n} - \sum_{\alpha}{\bf b}_{\alpha}\cdot{\rm d} 
{\bf v}_{s\alpha} ,
\label{Gibbs}
\end{equation}
which shows the pressure is a function of the intensive variables 
($T, \mu_{\alpha}, {\bf v}_{n}, {\bf v}_{s\alpha}$).  In analogy to 
binary mixtures of classical fluids, it is useful to introduce the
concentration variable 
\begin{equation} 
c = \rho_{1}/\rho, \,\,\,\,\,\, 1-c = \rho_{2}/\rho , 
\end{equation}
where $\rho = \rho_{1}+\rho_{2}$. If we define 
\begin{equation}
\mu \equiv \mu_{1}c + \mu_{2}(1-c), 
\end{equation}
then Eq.(\ref{ds}) can be written as 
\begin{equation}
T{\rm d}s = {\rm d}\epsilon
-{\bf v}_{n}\cdot {\rm d}{\bf g}
-\mu{\rm d}\rho - (\mu_{1} - \mu_{2})\rho{\rm d}c 
- \sum_{\alpha} {\bf b}_{\alpha}\cdot {\rm d}{\bf v}_{s\alpha}.
\end{equation}
Likewise, Eq.(\ref{pressure}) can be written as 
\begin{equation}
P =  Ts - \epsilon +\mu\rho + {\bf v}_{n}\cdot{\bf g},   
\end{equation}
while the Gibbs-Duham relation can be cast in the form
\begin{equation}
\delta P =  s\delta T + {\bf g}\cdot\delta {\bf v}_{n} + \rho\delta\mu 
- (\mu_{1} - \mu_{2})\rho\delta c 
- \sum_{\alpha}{\bf b}_{\alpha}\cdot\delta {\bf v}_{s\alpha}. 
\end{equation}

\vspace{0.2in}

\noindent \underline{\em The Galilean transformation properties of the
hydrodynamic variables} :
As we shall see, the derivation of the hydrodynamic equations can be simplified
by using the Galilean transformation (GT) properties of the hydrodynamic
variables $X_{i}$ and their derivatives 
$\partial S/\partial X_{i}$.  A Galilean transformation is a space-time
transformation $({\bf r}\rightarrow {\bf r}'={\bf r}+{\bf u}t$, $t\rightarrow
t'=t)$, where ${\bf u}$ is the relative velocity of the two inertial frames. 
Under this transformation, a scalar function $F$ of ${\bf r}$ and $t$
 will change to $F'$, which is a function of ${\bf r'}$ and $t'$. In 
particular, we have 
\begin{equation}
\rho'_{\alpha}=\rho_{\alpha},   \,\,\,\,\,\,  \alpha =1,2
 \,\,\,\,\,\,\,\,\,
\end{equation}
\begin{equation}
{\bf g}' = {\bf g} + \rho {\bf u}, \,\,\,\,\,\,   {\rm where} \,\,\,\,\,\,
 \rho = \rho_{1} +\rho_{2}, 
\label{gtrho} \end{equation}
\begin{equation}
\epsilon' = \epsilon + {\bf g}\cdot {\bf u} + \rho u^{2}/2,
\label{gtenergy} \end{equation}
and
\begin{equation}
{\bf v}_{s\alpha}' = {\bf v}_{s\alpha} + {\bf u}.
\label{gtvs} \end{equation}
While the validity of Eqs.(\ref{gtrho}) and (\ref{gtenergy}) is easy to see, 
Eq.(\ref{gtvs}) follows from the Galilean transformation 
properties of the field operator of the species $\alpha$, 
\begin{equation}
\hat{\psi}'_{\alpha}({\bf r'}, t') = \hat{\psi}_{\alpha}({\bf r}, t) 
{\rm exp}\left(\frac{im_{\alpha}{\bf u}\cdot{\bf r}}{\hbar}
+\frac{im_{\alpha}u^{2}t}{2\hbar}\right) . 
\end{equation}
The phase $\theta_{\alpha}$ of 
$<\hat{\psi}_{\alpha}({\bf r},t)>$ transforms as 
\begin{equation}
\theta'_{\alpha} = \theta_{\alpha} + 
\frac{m_{\alpha}{\bf u}\cdot {\bf r}}{\hbar} + 
\frac{m_{\alpha}u^{2}t}{2\hbar}, 
\end{equation}
hence Eq.(\ref{gtvs}), which says that ${\bf v}_{s}$ indeed transforms 
like a velocity field. Finally, we also note that the
momentum density of the species $\alpha$, written as ${\bf g}_\alpha$,
transforms as 
\begin{equation}
{\bf g}_{\alpha}\rightarrow {\bf g'}_{\alpha} = 
{\bf g}_{\alpha} + \rho_{\alpha} {\bf u}, \,\,\,\,\,\,\,\,\,\,
{\rm and} \,\,\,\,\,\,\,\,\,\,  {\bf g} = {\bf g}_{1} + {\bf g}_{2}.
\end{equation}

Since the entropy involves
counting of states, it is a Galilean invariant; we then have
\begin{equation}
s'\left( (\epsilon + {\bf g}\cdot{\bf u}+\frac{1}{2}\rho u^{2}),
({\bf g}+\rho {\bf u}), 
\rho_{1}, \rho_{2}, 
({\bf v}_{s1}+{\bf u}), ({\bf v}_{s2}+{\bf u})\right)
= s\left(\epsilon, {\bf g},
\rho_{1}, \rho_{2}, {\bf v}_{s1}, {\bf
v}_{s2}\right).
\label{sGT} \end{equation}
Differentiating both sides of Eq.(\ref{sGT}) with respect to $\epsilon$ 
and ${\bf v}_{s\alpha}$, we find that both $T$ and ${\bf b}_{\alpha}$ 
are Galilean invariant, 
\begin{equation} T'=T, \,\,\,\,\,\, {\bf b}_{\alpha}'= {\bf b}_{\alpha} . 
\label{srotation} \end{equation}
Differentiating both sides of 
Eq.(\ref{sGT}) with respect to ${\bf g}$ and $\rho_{\alpha}$, we get
\begin{equation}
{\bf v}_{n}' = {\bf v}_{n} + {\bf u}, \,\,\,\,\,\, {\rm and} 
\,\,\,\,\,\,
\mu'_{\alpha} = \mu_{\alpha}-{\bf v}_{n}\cdot{\bf u} -\frac{1}{2}u^{2}, 
\,\,\,\,\,\,\,\,\, \alpha = 1,2 .
\label{gtmu} \end{equation}
This shows that ${\bf v}_{n}$ also transforms like a velocity field. 
Finally, 
differentiating Eq.(\ref{sGT}) with respect to ${\bf u}$ at ${\bf u}=0$ gives 
\begin{equation}
{\bf b}_{1} + {\bf b}_{2} = {\bf g} - \rho {\bf v}_{n} \equiv {\bf g}^{0} , 
\label{go} \end{equation}
where the superscript $^{0}$ refers to quantities evaluated in the ${\bf v}_{n}=0$
frame. From Eq.(\ref{gtmu}), we then have 
\begin{equation}
\mu_{\alpha} = \mu_{\alpha}^{0} - \frac{1}{2} {\bf v}_{n}^{2} , 
\,\,\,\,\,\,\,\, \alpha = 1,2 ,
\end{equation}
and from Eqs.(\ref{gtrho}), (\ref{gtenergy}), and (\ref{gtmu}), 
it is easy to see that the pressure $P$ is a Galilean invariant and can
also be written as
\begin{equation}
P = Ts -\epsilon^{0} +\mu_{1}^{0}\rho_{1} + \mu_{2}^{0}\rho_{2}.
\label{PGT} \end{equation}

\noindent \underline{\em Rotational symmetry } : 
Under an infinitesimal rotation ${\bf a}$,  a vector such as 
 ${\bf g}$ changes to ${\bf g}+{\bf a}\times {\bf g}$. Since 
entropy of the system is invariant under such a rotation, we have 
\begin{equation}
s\left(\epsilon, ({\bf g} + {\bf a}\times{\bf g}),  \rho_{\alpha}, 
({\bf v}_{s1} + {\bf a}\times{\bf v}_{s1}), 
({\bf v}_{s2} + {\bf a}\times{\bf v}_{s2}) \right) 
= s(\epsilon, {\bf g}, \rho_{\alpha}, {\bf v}_{s1}, {\bf v}_{s2}).
\end{equation}
Differentiating both sides with respect to ${\bf a}$, we have 
\begin{equation}
{\bf v}_{n}\times {\bf g} + {\bf b}_{1}\times {\bf v}_{s1}
+ {\bf b}_{2}\times {\bf v}_{s2} = 0, 
\label{rot} \end{equation}
which implies ${\bf b}_{1}\times{\bf v}_{s1}^{0} + {\bf b}_{2}\times{\bf v}_{s2}^{0}=0$. 
\noindent The general form of ${\bf b}_{\alpha}$ satisfying Eq.(\ref{rot}) 
can be written as 
\begin{equation}
{\bf b}_{1} = \rho_{s1}{\bf v}_{s1}^{0} +
\tilde{\rho}({\bf v}_{s1}^{0}-{\bf v}_{s2}^{0}), 
\,\,\,\,\,\,\,
{\bf b}_{2} = \rho_{s2}{\bf v}_{s2}^{0} +
\tilde{\rho}({\bf v}_{s2}^{0}-{\bf v}_{s1}^{0}), \label{b1b2}
\end{equation}
where $\rho_{s1}, \rho_{s2}$, and $\tilde{\rho}$ are parameters which will 
be referred to as superfluid densities. 
With this parameterization, using Eq.(\ref{go}), we have 
\begin{equation}
{\bf g}^{0} = \rho_{s1}{\bf v}_{s1}^{0} + \rho_{s2}{\bf v}_{s2}^{0} \,\, , 
\end{equation}
\begin{equation}
{\bf g} =  \rho_{n}{\bf v}_{n} + \rho_{s1}{\bf v}_{s1} + 
\rho_{s2}{\bf v}_{s2}, \,\,\,\,\,\,\,\, \rho_{n} \equiv \rho -\rho_{s1} 
-\rho_{s2},
\end{equation}
where $\rho_{n}$ will be refered to as the normal fluid density. 
For later discussions, it is convenient to define the ``diffusion" current 
${\bf J}^{D}$ as 
\begin{equation}
{\bf g}^{0}_{1} \equiv  {\bf b}_{1} + {\bf J}^{D} , 
\,\,\,\,\,\,\,\,\,
{\bf g}^{0}_{2} \equiv  {\bf b}_{2} - {\bf J}^{D}. 
\label{gcrelation} \end{equation}
Using the above definition of ${\bf J}^{D}$, ${\bf g}_{\alpha}$ can be written as 
\begin{equation}
{\bf g}_1 = \rho_{n1}{\bf v}_{n}+ \rho_{s1}{\bf v}_{s1} 
+\tilde{\rho}({\bf v}_{s1}-{\bf v}_{s2}) + {\bf J}^{D} \,\, ,  \label{g1}
\end{equation}
\begin{equation}
{\bf g}_2 = \rho_{n2}{\bf v}_{n}+ \rho_{s2}{\bf v}_{s2} 
+ \tilde{\rho}({\bf v}_{s2}-{\bf v}_{s1}) - {\bf J}^{D},  \label{g2}
\end{equation}
where the normal density of the species $\alpha$ , $\rho_{n\alpha}$ is
defined as
\begin{equation}
\rho_{n\alpha} \equiv \rho_{\alpha} - \rho_{s\alpha}, 
\,\,\,\,\,\,\,  \rho_{n} = \rho_{n1} + \rho_{n2}.
\end{equation}
From Eqs.(\ref{g1}) and (\ref{g2}), it is clear that the $\tilde{\rho}$ term
describes the backflow created by one $superfluid$ 
component on the other in the mixture. 

\vspace{0.2in}

\noindent \underline{\em Continuity equations and symmetry relations } :
The time evolution of the hydrodynamic variables $\{ x_{i}\}$ are
governed by continuity equations and conservation laws, which are  

\begin{equation}
\partial_{t}\rho_{\alpha} = -{\bf \nabla}\cdot{\bf g}_{\alpha},
\,\,\,\,\,\, \alpha=1,2 , 
\label{rhocon} \end{equation}

\begin{equation}
\partial_{t}{\bf g}_{i} = -\rho_{1} {\bf \nabla}_{i}\phi_{1} -\rho_{2}
{\bf \nabla}_{i}\phi_{2} - \nabla_{j}\Pi_{ij},
\label{gcon} \end{equation}

\begin{equation}
\partial_{t}\epsilon = - {\bf \nabla}\cdot{\bf Q} -{\bf g}_{1}\cdot {\bf
\nabla}\phi_{1}  -{\bf g}_{2}\cdot {\bf \nabla}\phi_{2}, 
\label{energycon} \end{equation}

\begin{equation}
\partial_{t}{\bf v}_{s\alpha} = - {\bf \nabla}\chi_{\alpha}, \,\,\,\,\,\,
\alpha=1,2 , 
\label{vscon} \end{equation}
where $\chi_{\alpha}= - \frac{\hbar}{m}\partial_{t}\theta_{\alpha}$, and 
$\phi_{1}$ and $\phi_{2}$ are the external potential experienced by Bosons 
{\bf 1} and {\bf 2} respectively, Eqs.(\ref{rhocon})
to (\ref{energycon}) are continuity equations for density, momentum and energy
respectively. 
$\Pi_{ij}$ is the stress tensor and ${\bf Q}$ is the energy current. 
An argument of Martin {\em et al.,} shows that $\Pi_{ij}$ can always be 
taken to be a symmetric
tensor\cite{Martin}. The quantities 
$\{{\bf Q}, \Pi_{ij}, {\bf g}_{\alpha}, \chi_{\alpha} \} \equiv \{
{\cal J}_{i} \}$, referred to as "currents",  control the flow of the hydrodynamic variables 
$\{ x_{i} \}$. 

To complete the hydrodynamic description, we need to express the currents $\{
{\cal J}_{i} \}$ in terms of the hydrodynamic variables $\{ x_{i}\}$ so as to form a
closed set of equations. As explained in the next section, this can be done by
studying the entropy production of the system. The derivation can be simplified
considerably by using the Galilean transformation properties of $\{ 
{\cal J}_{i} \}$. By noting that the partial derivatives 
$\partial_{t}$ and ${\bf \nabla}$ transform as 
$\partial_{t} \rightarrow \partial_{t'} = \partial_{t} - {\bf u}\cdot {\bf
\nabla}$, and ${\bf \nabla} \rightarrow {\bf \nabla}' = {\bf \nabla}$ under a 
Galilean transformation, we have
\begin{equation}
Q'_i = Q_i + \Pi_{ij}u_j 
+ u_i\epsilon' + \frac{1}{2}(g_{1i} + 
g_{2i})u^2   \label{Q'}
\end{equation}
\begin{equation}
\Pi_{ij}' = \Pi_{ij} + u_i(g_1+ g_2)_j + (g_1+
g_2)_i u_j + \rho u_i u_j  \label{Pi'} 
\end{equation}
\begin{equation}
\chi_{\alpha}' = \chi_{\alpha} + {\bf u}\cdot{\bf v}_{s\alpha} +
\frac{u^{2}}{2}.
 \label{chi'}
\end{equation}
For example, Eq.(\ref{chi'}) is obtained by noting 
$\chi'_{\alpha} = - \frac{\hbar}{m_{\alpha}}
(\partial_{t} - {\bf u}\cdot{\bf \nabla})
(\theta_{\alpha} + m_{\alpha}{\bf u}\cdot{\bf r}/\hbar
 + m_{\alpha}u^{2}t/2\hbar)$
$=\chi_{\alpha} - u^{2}/2 +{\bf u}\cdot{\bf v}_{s\alpha} + u^{2}$. 

Using these properties, we can write Eqs.(\ref{Q'}) to (\ref{chi'}) as 
\begin{equation}
Q_i = Q^{0}_i + \Pi_{ij}^{0}v_{nj} 
+ v_{ni}\epsilon + \frac{1}{2}(g_1^{0} + g_2^{0})_i v_{n}^2 , \label{QGT}
\end{equation}
\begin{equation}
\Pi_{ij} = \Pi_{ij}^{0} + v_{ni}(g^{0}_1+ g^{0}_2)_j + 
(g^{0}_1+ g^{0}_2)_i v_{nj} + \rho v_{ni} v_{nj} , 
\label{PiGT} \end{equation}
\begin{equation}
\chi_{\alpha} = \chi_{\alpha}^{0} + {\bf v}_{n}\cdot{\bf v}_{s\alpha}^{0} +
\frac{1}{2}{\bf v}_{n}^2 .  \label{chiGT}
\end{equation}

For later use, we note that 
\begin{equation}
\left({\bf \nabla}\cdot{\bf g}\right)^{0} = {\bf \nabla}\cdot{\bf g}^{0}
+\rho {\bf \nabla}\cdot{\bf v}_{n} ,  \label{gradg0}
\end{equation}
\begin{equation}
({\bf \nabla}\cdot{\bf Q})^{0} = {\bf \nabla}\cdot{\bf Q}^{0} + 
\Pi^{0}_{ij} \partial_{j}v_{ni} + \epsilon^{0}({\bf \nabla}\cdot{\bf v}_{n}),
\label{gradQ0}
\end{equation}
\begin{equation}
({\bf \nabla}\chi_{\alpha})^{0}= {\bf \nabla}\chi_{\alpha}^{0} 
+  (v_{s\alpha}^{0})_{i}{\bf \nabla}v_{ni} .  \label{gradchi0}
\end{equation}

\vspace{0.2in}

\noindent {\bf II. Derivation  of the hydrodynamic equations for binary
mixtures of scalar superfluids : }

According to the second law of thermodynamics, the entropy production
of the system must be positive, $\frac{{\rm d}}{{\rm dt}}\int {\rm d}^{3}x 
s \geq 0$.  Note that 
\begin{equation}
\partial_{t}s = -\frac{1}{T}{\bf \nabla}\cdot {\bf Q} 
+ \frac{v_{ni}}{T}\partial_{j}\Pi_{ij}   
+ \sum_{\alpha=1,2}\left( -\frac{1}{T}{\bf g}_{\alpha}\cdot{\bf
\nabla}\phi_{\alpha}  + \frac{\mu_\alpha}{T}{\bf \nabla}\cdot{\bf g}_{\alpha}  
+\frac{1}{T}{\bf b}_{\alpha}\cdot{\nabla}\chi_{\alpha} +
\frac{1}{T}\rho_{\alpha}{\bf v_{n}}\cdot
{\bf \nabla}\phi_\alpha \right).
\label{entropypro} \end{equation}
Evaluating Eq.(\ref{entropypro}) in the ${\bf v}_{n}=0$ frame, we have
\begin{eqnarray}
T(\partial_ts)^0 &=& -\left({\bf \nabla}\cdot{\bf Q}^0 + 
\Pi_{ij}^0\partial_{i}v_{nj} + \epsilon^0{\bf \nabla}\cdot{\bf v}_{n}\right)  
+\nonumber\\
& & + \sum_{\alpha=1,2}[-{\bf g}_{\alpha}^{0}\cdot{\bf \nabla}\phi_\alpha
+ \mu_\alpha^0\rho_\alpha{\bf \nabla}\cdot{\bf v}_n + 
\mu_\alpha^0{\bf\nabla}\cdot{\bf g}_\alpha^0 +
{\bf b}_\alpha\cdot{\bf\nabla}\chi_\alpha^0 
\nonumber \\
& & + (v_{s\alpha}^0)_i({\bf b}_\alpha\cdot{\bf \nabla})v_{ni} ].
\label{enpro}
\end{eqnarray}
Using Eq.(\ref{PGT}), and Eqs.(\ref{gradg0}) to (\ref{gradchi0}), 
Eq.(\ref{enpro}) can be written as 
\begin{eqnarray}
T(\partial_ts)^0 &=& 
-{\bf \nabla}\cdot [{\bf Q}^0 + 
\sum_{\alpha=1,2}({\bf g}_{\alpha}^{0}\phi_\alpha -
{\bf b}_\alpha\chi_\alpha^0) ] +\nonumber\\
& &  -\left[\Pi_{ij}^0 + (Ts-P)\delta_{ij}
- \sum_{\alpha} (v_{s\alpha}^0)_{j}b_{\alpha i}\right] \partial_{i}v_{nj}
\nonumber\\
& & + \sum_{\alpha=1,2}[
(\phi_\alpha + \mu_{\alpha}^{0}){\bf \nabla}\cdot{\bf g}_{\alpha}^{0}
- \chi_\alpha^0{\bf\nabla}\cdot{\bf b}_\alpha ].
\label{newenpro}
\end{eqnarray}
By noting that 
$\partial_ts + {\bf \nabla}\cdot(s{\bf v}_{n})$$=$ 
$(\partial_ts)^0 + s{\bf \nabla}\cdot{\bf v}_{n}$ is Galilean invariant and 
using Eq.(\ref{gcrelation}), 
Eq.(\ref{newenpro}) can be written as 
\begin{equation}
T[\partial_ts + {\bf \nabla}\cdot(s{\bf v}_{n})] =
-{\bf \nabla}\cdot {\bf Q}^D -\Pi_{ij}^D\partial_{i}v_{nj} 
- \sum_{\alpha=1,2}\chi^{D}_{\alpha} {\bf \nabla}
\cdot {\bf b}_{\alpha} + \nu {\bf \nabla}\cdot {\bf J}^{D},
\label{fenpro}
\end{equation}
where ${\bf J}^{D}$ is defined in Eq.(\ref{gcrelation}). We introduce
$\Pi^{D}$, ${\bf Q}^{D}$, $\chi^{D}$, ${\bf J}^{D}$, and 
$\nu^{D}$ such that 
\begin{equation}
\Pi_{ij}^0 = \delta_{ij}P + \sum_{\alpha=1,2}
b_{\alpha i}(v_{s\alpha}^0)_{j} + \Pi_{ij}^D , \label{Pi0}
\end{equation}
\begin{equation}
{\bf Q}^0 = \sum_{\alpha=1,2}\left({\bf b}_\alpha \chi_\alpha^0 
-\phi_\alpha{\bf  g}_\alpha^0\right)  + {\bf Q}^D , \label{Q0}
\end{equation}
\begin{equation}
\chi_{\alpha}^0 = \phi_{\alpha} + \mu_{\alpha}^0 + 
\chi^{D}_{\alpha} , \label{chi0}
\end{equation}
\begin{equation}
\nu = (\mu_{1}^{0}+\phi_{1}) - (\mu_{2}^{0}+\phi_{2}) .  \label{nu}
\end{equation}
The determination of the ``currents" $\{ {\cal J}_{i} \}=\{ {\bf Q}, 
\Pi_{ij}, {\bf g}_{\alpha},\chi_{\alpha} \}$ now reduces to the 
determination of their ``dissipative" components $\{ {\cal J}_{i}^{D} \}
\equiv \{ {\bf Q}^{D}, \Pi_{ij}^{D},{\bf J}^{D}, 
\chi_{\alpha}^{D} \}$. 
Introducting $R_{1}$ and $R_{2}$ defined below, 
Eq.(\ref{entropypro}) becomes 
\begin{equation} 
\partial_ts +  {\bf \nabla}\cdot\left(s{\bf v}_n + \frac{{\bf Q}^D}{T}
- \frac{\nu {\bf J}^D}{T}\right) = R  \equiv R_{1} + R_{2},
\label{R} \end{equation}
\begin{equation}
R_{1}  = -\frac{1}{T}\left[\chi^{D}_1{\bf
 \nabla}\cdot{\bf b}_1 + \chi^{D}_2{\bf\nabla}\cdot{\bf b}_2 +
\Pi_{ij}^D\partial_i v_{nj}\right] \,\,\, , 
\label{R1} 
\end{equation}
\begin{equation}
R_{2} = {\bf Q}^D\cdot{\bf \nabla}\left(\frac{1}{T}\right) -
{\bf J}^D\cdot{\bf \nabla}\left(\frac{\nu}{T}\right) 
= - ({\bf Q}^D - \nu {\bf J}^D)\cdot \left( \frac{ {\bf \nabla}T}{T^2} \right)
- {\bf J}^D\cdot \left( \frac{{\bf \nabla}\nu}{T}\right).
\label{R2} \end{equation}
The right hand side 
of Eq.(\ref{R}) expresses the entropy generated  at ${\bf r}$ by the
``external forces" $\{ f_{i} \}\equiv \{ -{\bf \nabla}T, 
\partial_{i}v_{nj},  -{\bf \nabla}\nu, {\bf \nabla}\cdot {\bf b}_{\alpha}\} $. The 
dissipative ``currents"  $\{ {\cal J}_{i}^{D} \}$ are responses 
to the disturbances $\{ f_{i} \}$ as the system tries to relax 
to equilibrium. In the linear response limit,  $\{ {\cal J}_{i}^{D} \}$ is
linear in $\{ f_{i} \}$, and 
$R$ becomes a quadratic form in the forces, which must be positive in order to 
satisfy the second law of thermodynamics. 
Once the relations between the dissipative currents  $\{ {\cal J}_{i}^{D} \}$
 and the driving forces $\{ f_{i} \}$ are determined, 
the hydrodynamic equations, given by Eqs.(\ref{rhocon}) to (\ref{vscon}) will form a 
closed set.  
Using Eqs.(\ref{g1}) and (\ref{g2}) for ${\bf g}_{\alpha}$, Eqs.(\ref{PiGT})
and (\ref{Pi0}) for $\Pi_{ij}$, Eqs.(\ref{chiGT}) and (\ref{chi0}) for
$\chi_{\alpha}$, and Eqs.(\ref{QGT}) and (\ref{Q0}) for ${\bf Q}$, we can write 
Eqs.(\ref{rhocon}) to (\ref{vscon}) as 
\begin{eqnarray}
{\bf I.} \hspace{0.2in} &
\partial_{t}\rho_{\alpha} = -{\bf \nabla}\cdot{\bf g}_{\alpha},
\,\,\,\,\,\, \alpha=1,2,  \label{I} \\
 & {\bf g}_1 = \rho_{n1}{\bf v}_{n} + \rho_{s1}{\bf v}_{s1} +
\tilde{\rho} \left({\bf v}_{s1}-{\bf v}_{s2}\right) + {\bf J}^D, \\
 & {\bf g}_2 = \rho_{n2}{\bf v}_{n} + \rho_{s2}{\bf v}_{s2} +
\tilde{\rho} \left({\bf v}_{s2}-{\bf v}_{s1}\right) - {\bf J}^D, \\
{\bf II.} \hspace{0.2in} & 
\partial_{t}({\bf g}_{1} + {\bf g}_{2})_{i} = -\rho_{1} \nabla_{i}\phi_{1}
-\rho_{2}\nabla_{i}\phi_{2} - \nabla_{j}(\Pi_{ij}^{r}+\Pi_{ij}^{D}), 
\label{II} \\
 & \Pi^{r}_{ij} = \delta_{ij}P + 
\left(\rho_{n1} + \rho_{n2}\right) v_{ni}v_{nj} + 
\sum_{\alpha=1,2}\rho_{s\alpha}\left(v_{s\alpha}\right)_i
\left(v_{s\alpha}\right)_j 
\nonumber \\
&  + \tilde{\rho}\left(v_{s1}-v_{s2}\right)_i \left(v_{s1}-v_{s2}\right)_j,
\end{eqnarray}
\begin{equation}
{\bf III.} \hspace{0.2in}
\partial_{t}{\bf v}_{s\alpha} = -{\bf \nabla}\left( \phi_{\alpha}+ \mu_{\alpha}
+{\bf v}_{n}\cdot{\bf v}_{s\alpha} +\chi^{D}\right), \,\,\,\,\,\,
\alpha=1,2 ,
 \label{III} \end{equation}
\begin{equation}
{\bf IV.} \hspace{0.2in}
\partial_ts +  {\bf \nabla}\cdot\left(s{\bf v}_n\right) 
+ {\bf \nabla}\cdot\left(\frac{{\bf Q}^D}{T} 
- \frac{\nu {\bf J}^D}{T}\right) = R,  \label{entropyeq}
\end{equation}
where we have replaced the energy equation Eq.(\ref{energycon}) by the entropy
equation Eq.(\ref{entropyeq}). 
For hydrodynamic processes with weak dissipation, 
the dissipative currents ${\bf J}^{D}, \Pi^{D}, \chi^{D}, {\bf Q}^{D}$ 
can be ignored and the resulting equations {\bf I} to {\bf IV} above
will be referred to as  ``dissipation-free" hydrodynamic equations. 
For dissipative processes, however, it is necessary to
determine the explicit form of the dissipative currents. 

\vspace{0,2in}

\noindent \underline{ Determination of dissipative currents } :
Note that the forces in $R_{1}$ and $R_{2}$ (eq.(\ref{R1}) and Eq.(\ref{R2}) )
 have different parity and time reversal symmetries. As a result, the currents
$( \chi_{\alpha}, \Pi^{D}_{ij} )$ and $( {\bf Q}^{D}, {\bf J}^{D} )$ are only 
linear combinations of $( {\bf \nabla}\cdot 
{\bf b}_{\alpha}, \partial_{i}v_{nj} 
)$ and $( {\bf \nabla}T, {\nabla}\nu )$ respectively. 
Since $\Pi_{ij}^D$ is symmetric, it can be decomposed as 
\begin{equation}
\Pi_{ij}^D = A_{ij} + \tau \delta_{ij} , \,\,\,\,\,\, {\rm Tr}A = 0, 
\end{equation}
which allows $R_1$ to be expressed as 
\begin{equation} 
R_{1} = -\frac{1}{T} \left[ \frac{1}{2}A_{ij}\left( 
\nabla_i v_{nj} + \nabla_j  v_{ni} -
\frac{2}{3}\delta_{ij}{\bf\nabla}\cdot{\bf v}_n \right)  
+ \tau {\bf \nabla}\cdot{\bf v}_{n} + 
\sum_{\alpha = 1,2} \chi_{\alpha} ({\bf \nabla}\cdot {\bf b}_{\alpha})
\right]. 
\end{equation}
Linear response implies 
\begin{equation}
A_{ij} = -\eta \left( \nabla_i v_{nj} + \nabla_j v_{ni} -
\frac{2}{3}\delta_{ij}{\bf\nabla}\cdot{\bf v}_n \right),
\end{equation}
\begin{eqnarray}
 \left( \begin{array}{c}
          \tau\\ \chi^{D}_1\\ \chi^{D}_2
          \end{array} \right)
    = - \left( \begin{array}{ccc} 
                              \zeta_{00} & \zeta_{01}  & \zeta_{02} \\
                              \zeta_{01} & \zeta_{11}  & \zeta_{12} \\
                              \zeta_{02} & \zeta_{21}  & \zeta_{22}
                              \end{array} \right)   
      \left( \begin{array}{c}
       {\bf\nabla}\cdot{\bf v}_n \\ {\bf\nabla}\cdot{\bf b}_1 \\
{\bf\nabla}\cdot{\bf b}_2
           \end{array} \right),             \label{tau}
\end{eqnarray}
where $\eta$ will be referred to as the ``first" viscosity, and 
$\zeta_{ij}$ as the ``second" viscosity matrix.  
The fact that $\zeta_{ij}$ is symmetric is a consequence of the 
general Onsager relations for kinetic coefficients. 
More explicitly, we have
\begin{equation}
\Pi_{ij}^D = - \eta \left(
\nabla_i v_{nj} + \nabla_j  v_{ni} -
\frac{2}{3}\delta_{ij}{\bf\nabla}\cdot{\bf v}_n \right) 
- \zeta_{00}{\bf \nabla}\cdot{\bf v}_{n} 
- \zeta_{01}{\bf \nabla}\cdot{\bf b}_{1}
- \zeta_{02}{\bf \nabla}\cdot{\bf b}_{2},
\label{PiDfinal} \end{equation}
and
\begin{equation}
\chi^{D}_{\alpha} = - \zeta_{0\alpha}{\bf \nabla}\cdot{\bf v}_{n} 
- \sum_{\sigma=1,2} \zeta_{\alpha \sigma}{\bf \nabla}\cdot{\bf b}_{\sigma}.
\label{chifinal}\end{equation}
Recall from Eq.(\ref{b1b2}) that
${\bf b}_{1} = \rho_{s1}{\bf v}_{s1}^{0} +
\tilde{\rho}({\bf v}_{s1}^{0}-{\bf v}_{s2}^{0})$ and 
${\bf b}_{2} = \rho_{s2}{\bf v}_{s2}^{0} +
\tilde{\rho}({\bf v}_{s2}^{0}-{\bf v}_{s1}^{0}).$
Clearly, in order to have $R_{1} \geq 0$, we need to have $\eta >0$, 
and det$|\zeta| \geq 0$. 

To find the dissipative currents ${\bf Q}^D$ and ${\bf J}^D$, we note that the
expression for $R_{2}$ is identical to the entropy production of a classical
binary mixture\cite{Landaufluid}. The expressions of these currents are 
therefore identical to those of classical fluids and are of the form 
\begin{equation}
{\bf J}^{D} = -\alpha T \left( \frac{ {\bf\nabla}\nu}{T}\right)  - \beta T^2 
\left(\frac{ {\bf\nabla}T}{T^2}\right)
\label{JD} \end{equation}
\begin{equation}
{\bf Q}^{D} - \nu {\bf J}^{D}  = -\beta T^2 \left( \frac{
{\bf\nabla}\nu}{T}\right) - \gamma T^2 \left(\frac{ {\bf\nabla}T}{T^2}\right) ,
\label{QJD} \end{equation}
where we have again used the Onsager relation to identify the off-diagonal 
kinetic coefficients in Eqs.(\ref{JD}) and (\ref{QJD}). 
Eq.(\ref{R2}) then becomes 
$R_{2} = \gamma({\bf \nabla}T)^2/T^2 + 
\alpha({\bf \nabla}\nu)^2/T + 
2\beta\left({\bf \nabla}T\cdot {\bf \nabla}\nu\right)/T$, or, 
\begin{equation}
R_{2} = \kappa \frac{({\bf \nabla}T)^2}{T^2} + \frac{1}{\alpha T}
\left({\bf J}^{D}\right)^2,
\end{equation}
where $\kappa \equiv \gamma - \beta^2 T/\alpha$. The positivity of $R_{2}$
implies $\kappa \geq 0$, and $\alpha \geq 0$. 
Eqs.(\ref{PiDfinal}), (\ref{chifinal}), 
(\ref{JD}), and (\ref{QJD}) completely determine the dissipative currents
and complete the full dissipative hydrodynamic equations, given by 
Eqs.(\ref{I}) to (\ref{entropyeq}) together with Eqs.(\ref{PiDfinal}) to 
and Eq.(\ref{QJD}). 

While we shall not be solving the hydrodynamic equations here, we remind the 
readers that the sound modes of the system are contained in 
the so-called linearized hydrodynamic equations, which are obtained 
by linearizing  Eqs.(\ref{I}) to (\ref{entropyeq}) about the equilibrium 
configuration. In the single component superfluid case, Landau\cite{Landau}
has pointed out that it is useful to express all the hydrodynamic 
variables in terms of externally controllable quantities such as 
temperature $T$ and pressure $P$.
(Also see reference \cite{VHsound}).  In the 
mixture case, there is another controllable quantity, which is the
concentration $c=\rho_{1}/\rho$ defined earlier in Section {\bf I}. 
Expressing ${\bf J}^D$ and ${\bf Q}^D$ in terms of $T, P$ and $c$ as in 
Landau and Lifshitz\cite{Landaufluid}, we have
\begin{equation}
{\bf J}^{D} = - \rho D \left({\bf \nabla}c + \frac{k_T}{T}{\bf\nabla}T
+ \frac{k_P}{P}{\bf\nabla}p\right), 
\label{dcurrent} \end{equation}
\begin{equation}
{\bf Q}^{D} = \left[ k_T\left(\frac{\partial \nu}{\partial c}\right)_{P,T}
- T\left(\frac{\partial \nu}{\partial T}\right)_{P,c} + \nu \right] {\bf J}^D
 - \kappa{\bf \nabla}T,
\label{hcurrent} \end{equation}
where $\kappa, D, k_{T}, k_{P}$ are the thermal conductivity, 
the diffusion coefficient, the coefficient of thermal diffusion, and 
the coefficient of baro-diffusion respectively; they can be written as 
\begin{equation}
\kappa = \gamma - \beta^2 T/\alpha, \,\,\,\,\,\,\,\,\,
D = \frac{\alpha}{\rho}\left(\frac{\partial \nu}{\partial
c}\right)_{T,P} \end{equation}
\begin{equation}
\rho D k_{T}/T  = \alpha\left(\frac{\partial \nu}{\partial
T}\right)_{P,c} + \beta , \,\,\,\,\,\,\,\,\,
\rho D k_{P}/P = \alpha \left( \frac{\partial \nu}{\partial P}\right)_{T, c}. 
\end{equation}
Using the relations above, the full expression for entropy production is 
\begin{eqnarray}
R &=& \frac{\eta}{2T}\left( 
\nabla_i v_{nj} + \nabla_j v_{ni} -
\frac{2}{3}\delta_{ij}{\bf\nabla}\cdot{\bf v}_n \right)^2  \nonumber \\
& & 
+ \frac{\zeta_{00}}{T}\left({\bf \nabla}\cdot{\bf  v}_n\right)^2 
+ 2\sum_{\alpha=1,2}\frac{\zeta_{0\alpha}}{T}
\left({\bf \nabla}\cdot{\bf v}_n\right)
\left({\bf \nabla}\cdot{\bf b}_\alpha\right) \nonumber \\
& &
+ \sum_{\alpha=1,2}\frac{\zeta_{\alpha\alpha}}{T}
\left({\bf \nabla}\cdot{\bf b}_\alpha\right)^2 +
+ 2\sum_{\alpha, \sigma=1,2}\frac{\zeta_{\alpha\sigma}}{T}
\left({\bf \nabla}\cdot{\bf b}_\alpha\right)
\left({\bf \nabla}\cdot{\bf b}_\sigma\right) \\
\nonumber \\
& & + \kappa \frac{\left({\bf \nabla}T\right)^2}{T^2} + 
\frac{D\rho}{T} \frac{\partial \nu}{\partial c}
\left({\bf \nabla}c + \frac{\kappa_T}{T}{\bf \nabla}T
+ \frac{\kappa_P}{P}{\bf \nabla}p\right)^2 . 
\end{eqnarray}

\vspace{0.2IN}

\noindent {\bf III.  Hydrodynamic equations of other mixture systems :}

\noindent (a) \underline{ Binary mixture of a normal fluid and a superfluid
} :  Let component 
{\bf 1} be the superfluid and component {\bf 2} be the normal fluid. Since 
{\bf 2} does not have broken gauge symmetry, ${\bf v}_{s2}$ is absent in the
hydrodynamic equations.  Rotational symmetry  
(Eq.(\ref{srotation})) yeilds 
${\bf b}_{1}\times {\bf v}_{s1}^{o} =0$, which implies 
${\bf b}_{1} = \rho_{s1} {\bf v}_{s1}$, or $\rho_{2}=\tilde{\rho} =0$. 
Following exactly the same steps as before, it is easy to see that the 
the hydrodynamic equations of this system are still given by Eqs.(\ref{I}) 
to (\ref{entropyeq}) with the quantities ($\rho_{s2}, \tilde{\rho}, 
{\bf v}_{s2}$) all
set to zero. Moreover, the dissipative part of the stress tensor
$\Pi^{D}_{ij}$ as well as $\chi^{D}_{1}$ are given by 
Eqs.(\ref{PiDfinal}) and (\ref{chifinal}) with $\zeta_{02}$ and ${\bf b}_{2}$
 set to zero, while the 
expressions of the diffusion current ${\bf J}^{D}$ and the heat current 
${\bf Q}^D$ are still given by Eqs.(\ref{dcurrent}) and (\ref{hcurrent}). 

\vspace{0.2in}

\noindent (b)\underline{ Binary mixture of normal fluids } : 

In this case, both ${\bf v}_{s1}$ and ${\bf v}_{s2}$ are absent from the
equations. The hydrodynamic equations are given by Eqs.(\ref{I}) to 
Eq.(\ref{entropyeq}) with the quantities 
($\rho_{s1}, \rho_{s2}, \tilde{\rho}, {\bf v}_{s1}, {\bf v}_{s2}$) set to
zero. 
The dissipative component $\Pi^D$ is given by
Eq.(\ref{PiDfinal}) with all $\zeta_{\alpha\beta}=0$
except for $\zeta_{00}$, and $\chi^D=0$. 
The diffusion current and heat current are still 
given by Eqs.(\ref{dcurrent}) and (\ref{hcurrent}). 

\vspace{0.2in}

Acknowledgement : This work is support in part by NSF Grant DMR-9705295.

\end{document}